\documentclass[11pt,a4paper]{article}

\AtBeginDocument{}
\newif\ifmetiszh
\metiszhfalse
\usepackage{metis-arxiv}
\newcommand{\figtext}[2]{\ifmetiszh #2\else #1\fi}

\tikzset{
  metisbox/.style={
    draw=MetisRule,
    rounded corners=2pt,
    line width=0.5pt,
    align=center,
    inner sep=4pt,
    font=\sffamily\scriptsize,
    text=MetisInk
  },
  metisarrow/.style={-{Latex[length=2.2mm]},line width=0.65pt,color=MetisGray},
  metisdash/.style={-{Latex[length=2.2mm]},line width=0.6pt,dashed,color=MetisGray}
}

\newcommand{\MetisArchitectureFigure}{%
\begin{tikzpicture}[x=1cm,y=1cm,font=\sffamily\tiny]
  \tikzset{
    archlabel/.style={
      draw=MetisRule,rounded corners=2pt,fill=MetisNavy,
      text=white,text width=13mm,minimum height=10mm,
      align=center,font=\sffamily\tiny\bfseries,inner sep=2pt
    },
    archbox/.style={
      draw=MetisRule,rounded corners=2pt,line width=0.45pt,
      align=center,minimum height=10mm,inner sep=2.5pt,
      font=\sffamily\tiny,text=MetisInk
    }
  }

  \node[font=\sffamily\scriptsize\bfseries,text=MetisNavy] at (8.15,8.32)
    {\figtext{METIS SYSTEM ARCHITECTURE}{METIS 系统架构}};

  \node[archlabel] (l-access) at (0.78,7.36)
    {\figtext{Access}{接入层}};
  \node[archbox,fill=MetisSky,text width=27mm] (tui) at (3.10,7.36)
    {\textbf{\figtext{Interactive CLI / TUI}{交互式 CLI / TUI}}\\
     chat, slash, permission UI};
  \node[archbox,fill=MetisSky,text width=27mm] (headless) at (6.45,7.36)
    {\textbf{\figtext{Headless execution}{无界面执行}}\\
     run / daemon / cron / eval};
  \node[archbox,fill=MetisSky,text width=27mm] (protocols) at (9.80,7.36)
    {\textbf{\figtext{Protocol servers}{协议服务}}\\
     ACP / MCP serve / coordinator};
  \node[archbox,fill=MetisSky,text width=27mm] (desktop) at (13.15,7.36)
    {\textbf{\figtext{Desktop and IDE}{桌面端与 IDE}}\\
     Wails / WebUI / IDE MCP};

  \node[archlabel] (l-runtime) at (0.78,5.65)
    {\figtext{Runtime}{运行时层}};
  \node[archbox,fill=MetisSky,text width=35mm] (providers) at (3.55,5.65)
    {\textbf{\figtext{Provider router and adapters}{Provider 路由与适配器}}\\
     Anthropic / OpenAI / Gemini / custom\\
     stream normalization + usage};
  \node[archbox,fill=MetisMint,text width=38mm] (loop) at (8.00,5.65)
    {\textbf{\figtext{Stateful agent loop}{有状态 Agent 循环}}\\
     request $\rightarrow$ stream $\rightarrow$ assistant blocks\\
     typed events + ordered iteration};
  \node[archbox,fill=MetisPanel,text width=35mm] (state) at (12.55,5.65)
    {\textbf{\figtext{History and lifecycle state}{历史与生命周期状态}}\\
     context tiers / memory / budget\\
     session / checkpoint / spill};

  \node[archlabel] (l-control) at (0.78,3.86)
    {\figtext{Control}{控制层}};
  \node[archbox,fill=MetisSand,text width=35mm] (registry) at (3.55,3.86)
    {\textbf{\figtext{Capability resolution}{能力解析}}\\
     registry / aliases / enabled checks\\
     schemas / lazy ToolSearch};
  \node[archbox,fill=MetisSand,text width=38mm] (gate) at (8.00,3.86)
    {\textbf{\figtext{Permission and hook plane}{权限与 Hook 平面}}\\
     five modes / rules / path scope\\
     pre-tool veto + batch ASK};
  \node[archbox,fill=MetisSand,text width=35mm] (dispatch) at (12.55,3.86)
    {\textbf{\figtext{Typed execution control}{类型化执行控制}}\\
     Safe / Queue / Exclusive / Background\\
     result closure + orphan repair};

  \node[archlabel] (l-ext) at (0.78,2.07)
    {\figtext{Extensions}{扩展层}};
  \node[archbox,fill=MetisMint,text width=27mm] (builtin) at (3.10,2.07)
    {\textbf{\figtext{Built-ins}{内置能力}}\\
     files / shell / git / web / Workflow};
  \node[archbox,fill=MetisMint,text width=27mm] (skill) at (6.45,2.07)
    {\textbf{\figtext{Skills and plugins}{Skill 与插件}}\\
     layered loader / public SDK};
  \node[archbox,fill=MetisMint,text width=27mm] (mcp) at (9.80,2.07)
    {\textbf{\figtext{MCP clients}{MCP 客户端}}\\
     lazy servers / resources / prompts};
  \node[archbox,fill=MetisMint,text width=27mm] (children) at (13.15,2.07)
    {\textbf{\figtext{Child execution}{子执行单元}}\\
     agents / roster / tasks / jobs\\
     worktree + transcript};

  \node[archlabel] (l-effect) at (0.78,0.30)
    {\figtext{Effects}{副作用层}};
  \node[archbox,fill=MetisRose,text width=27mm] (localfx) at (3.10,0.30)
    {\textbf{\figtext{Local workspace}{本地工作区}}\\
     file / process / Git / network};
  \node[archbox,fill=MetisRose,text width=27mm] (stores) at (6.45,0.30)
    {\textbf{\figtext{Persistent stores}{持久化存储}}\\
     sessions / memory / jobs / cache};
  \node[archbox,fill=MetisRose,text width=27mm] (external) at (9.80,0.30)
    {\textbf{\figtext{External services}{外部服务}}\\
     MCP / APIs / channels};
  \node[archbox,fill=MetisRose,text width=27mm] (cu) at (13.15,0.30)
    {\textbf{\figtext{Companion service}{配套服务}}\\
     MCP gate $\rightarrow$ frontmost app\\
     display / pointer / keyboard};

  \coordinate (access-bus-left) at ($(tui.south)+(0,-0.14)$);
  \coordinate (access-bus-right) at ($(desktop.south)+(0,-0.14)$);
  \draw[line width=0.45pt,color=MetisGray]
    (access-bus-left) -- (access-bus-right);
  \foreach \a in {tui,headless,protocols,desktop}
    \draw[line width=0.45pt,color=MetisGray]
      (\a.south) -- (\a.south |- access-bus-left);
  \draw[metisarrow] (loop.north |- access-bus-left) -- (loop.north);
  \draw[metisarrow] (providers.east) -- (loop.west);
  \draw[metisarrow] (state.west) -- (loop.east);
  \draw[metisarrow] (loop.south) -- ++(0,-0.25) -| (registry.north);
  \draw[metisarrow] (registry.east) -- (gate.west);
  \draw[metisarrow] (gate.east) -- (dispatch.west);
  \coordinate (extension-bus-left) at ($(builtin.north)+(0,0.24)$);
  \coordinate (extension-bus-right) at ($(children.north)+(0,0.24)$);
  \draw[line width=0.45pt,color=MetisGray]
    (extension-bus-left) -- (extension-bus-right);
  \draw[metisarrow] (dispatch.south) -- (dispatch.south |- extension-bus-left);
  \foreach \a in {builtin,skill,mcp,children}
    \draw[metisarrow] (\a.north |- extension-bus-left) -- (\a.north);
  \draw[metisarrow] (builtin.south) -- (localfx.north);
  \draw[metisarrow] (skill.south) -- (stores.north);
  \draw[metisarrow] (mcp.south) -- (external.north);
  \draw[metisarrow] (children.south) -- (cu.north);

  \begin{scope}[on background layer]
    \node[draw=MetisRule,rounded corners=5pt,fit=(l-access)(desktop)(l-effect)(cu),
      inner sep=4pt] {};
    \node[draw=MetisBlue,dashed,rounded corners=4pt,fit=(providers),
      inner sep=3pt,label={[font=\sffamily\tiny,text=MetisBlue,fill=white,inner sep=1pt]above:
      \figtext{model/API boundary}{模型/API 边界}}] {};
    \node[draw=MetisTeal,dashed,rounded corners=4pt,
      fit=(loop)(state)(gate)(registry)(dispatch)(builtin)(skill)(mcp)(children),
      inner sep=4pt,label={[font=\sffamily\tiny,text=MetisTeal,fill=white,inner sep=1pt]below:
      \figtext{Metis process boundary}{Metis 进程边界}}] {};
    \node[draw=MetisAmber,dashed,rounded corners=4pt,
      fit=(localfx)(stores)(external)(cu),
      inner sep=3pt,label={[font=\sffamily\tiny,text=MetisAmber,fill=white,inner sep=1pt]below:
      \figtext{effect and persistence boundary}{副作用与持久化边界}}] {};
  \end{scope}
\end{tikzpicture}%
}

\newcommand{\DispatchTimelineFigure}{%
\begin{tikzpicture}[x=0.073cm,y=0.62cm,font=\sffamily\tiny]
  \draw[->,color=MetisGray] (0,-0.15) -- (76,-0.15)
    node[right]{\figtext{time}{时间}};
  \foreach \x in {0,30,60,70} {
    \draw[color=MetisRule] (\x,-0.28) -- (\x,5.45);
    \node[below,text=MetisGray] at (\x,-0.28) {\x};
  }
  \node[anchor=east] at (-2,4.8) {Safe 1};
  \node[anchor=east] at (-2,4.0) {Safe 2};
  \node[anchor=east] at (-2,3.2) {Queue 1};
  \node[anchor=east] at (-2,2.4) {Queue 2};
  \node[anchor=east] at (-2,1.6) {Background};
  \node[anchor=east] at (-2,0.8) {Exclusive};

  \fill[MetisBlue] (0,4.52) rectangle (30,5.08);
  \fill[MetisBlue] (0,3.72) rectangle (30,4.28);
  \fill[MetisTeal] (0,2.92) rectangle (30,3.48);
  \fill[MetisTeal!65] (30,2.12) rectangle (60,2.68);
  \fill[MetisAmber] (0,1.32) rectangle (5,1.88);
  \draw[MetisAmber,dashed,line width=1pt] (5,1.6) -- (70,1.6);
  \fill[MetisRed] (60,0.52) rectangle (70,1.08);

  \node[text=white] at (15,4.8) {\figtext{fan-out}{并发}};
  \node[text=white] at (15,4.0) {\figtext{fan-out}{并发}};
  \node[text=white] at (15,3.2) {FIFO};
  \node[text=white] at (45,2.4) {FIFO};
  \node[anchor=west,text=MetisAmber] at (6,1.86) {\figtext{detached work}{脱离式任务}};
  \node[text=white] at (65,0.8) {\figtext{barrier}{屏障}};

  \draw[decorate,decoration={brace,amplitude=3pt},MetisGray]
    (0,5.28) -- (60,5.28)
    node[midway,above=4pt]{\figtext{Safe and Queue overlap; Queue remains serial}{Safe 与 Queue 重叠，Queue 内部串行}};
\end{tikzpicture}%
}

\newcommand{\HistoryClosureFigure}{%
\begin{tikzpicture}[node distance=4mm and 5mm]
  \node[metisbox,fill=MetisSky,text width=15mm] (u1) {\code{use(a)}};
  \node[metisbox,fill=MetisMint,text width=16mm,right=of u1] (r1) {\code{result(a)}};
  \node[metisbox,fill=MetisSky,text width=15mm,right=of r1] (u2) {\code{use(b)}};
  \node[metisbox,fill=MetisPanel,text width=16mm,right=of u2] (text) {\figtext{user text}{用户文本}};
  \node[metisbox,fill=MetisRose,text width=17mm,right=of text] (bad) {\figtext{missing}{缺失}\\\code{result(b)}};

  \draw[metisarrow] (u1) -- (r1);
  \draw[metisarrow] (r1) -- (u2);
  \draw[metisarrow] (u2) -- (text);
  \draw[metisdash] (text) -- (bad);
  \draw[MetisRed,line width=0.8pt] ($(u2.north east)+(1mm,3mm)$) -- ($(u2.south east)+(1mm,-3mm)$);
  \node[font=\sffamily\tiny,text=MetisRed,above=5mm of u2.east]
    {\figtext{unsafe cut}{不安全切点}};

  \node[metisbox,fill=MetisSky,text width=15mm,below=10mm of u1] (ou1) {\code{use(a)}};
  \node[metisbox,fill=MetisMint,text width=16mm,right=of ou1] (or1) {\code{result(a)}};
  \node[metisbox,fill=MetisSky,text width=15mm,right=of or1] (ou2) {\code{use(b)}};
  \node[metisbox,fill=MetisRose,text width=22mm,right=of ou2] (stub)
    {\code{result(b)}\\[-1pt]\figtext{interrupted}{已中断}};
  \draw[metisarrow] (ou1) -- (or1);
  \draw[metisarrow] (or1) -- (ou2);
  \draw[metisarrow] (ou2) -- (stub);
  \draw[metisdash,draw=MetisAmber] (bad.south) to[bend left=12]
    node[right,font=\sffamily\tiny,text=MetisAmber]
    {\figtext{idempotent repair}{幂等修复}} (stub.north);

  \node[font=\sffamily\scriptsize\bfseries,text=MetisNavy,above=2mm of u1]
    {\figtext{Observed history}{原历史}};
  \node[font=\sffamily\scriptsize\bfseries,text=MetisNavy,below=2mm of ou1]
    {\figtext{Provider-valid continuation}{Provider 可接受的续接}};
\end{tikzpicture}%
}

\newcommand{\RuntimeEventGraphFigure}{%
\begin{tikzpicture}[node distance=4mm and 5mm]
  \node[metisbox,fill=MetisSky,text width=20mm,minimum height=11mm] (wire)
    {\textbf{\figtext{Provider wire}{Provider 传输层}}\\[-1pt]\scriptsize deltas / stop\\[-1pt]/ usage};
  \node[metisbox,fill=MetisSky,text width=22mm,minimum height=11mm,right=of wire] (normal)
    {\textbf{\figtext{Normalization}{归一化}}\\[-1pt]\scriptsize text / thinking\\[-1pt]/ args};
  \node[metisbox,fill=MetisMint,text width=22mm,minimum height=11mm,right=of normal] (assistant)
    {\textbf{\figtext{Assistant blocks}{Assistant 内容块}}\\[-1pt]\scriptsize ordered\\[-1pt]tool uses};
  \node[metisbox,fill=MetisSand,text width=20mm,minimum height=11mm,right=of assistant] (permit)
    {\textbf{\figtext{Permission}{权限}}\\[-1pt]\scriptsize allow / ask\\[-1pt]/ deny};
  \node[metisbox,fill=MetisSand,text width=21mm,minimum height=11mm,right=of permit] (exec)
    {\textbf{\figtext{Execution}{执行}}\\[-1pt]\scriptsize class + hooks};
  \node[metisbox,fill=MetisRose,text width=20mm,minimum height=11mm,right=of exec] (result)
    {\textbf{\figtext{Terminal result}{终止结果}}\\[-1pt]\scriptsize value / error\\[-1pt]/ stub};

  \draw[metisarrow] (wire) -- (normal);
  \draw[metisarrow] (normal) -- (assistant);
  \draw[metisarrow] (assistant) -- (permit);
  \draw[metisarrow] (permit) -- (exec);
  \draw[metisarrow] (exec) -- (result);

  \node[metisbox,fill=MetisPanel,text width=30mm,below=9mm of normal] (telemetry)
    {\figtext{Observable stream events}{可观测流事件}\\[-1pt]\scriptsize text / thinking / args\\[-1pt]tokens / stop};
  \node[metisbox,fill=MetisPanel,text width=30mm,below=9mm of permit] (control)
    {\figtext{Control and safety events}{控制与安全事件}\\[-1pt]\scriptsize permission / hook\\[-1pt]cancel / panic};
  \node[metisbox,fill=MetisPanel,text width=31mm,below=9mm of result] (lifecycle)
    {\figtext{Lifecycle events}{生命周期事件}\\[-1pt]\scriptsize compact / sub-agent\\[-1pt]job / fallback};
  \draw[metisdash] (normal) -- (telemetry);
  \draw[metisdash] (permit) -- (control);
  \draw[metisdash] (result) -- (lifecycle);
  \draw[metisarrow] (result.south west) to[bend left=16]
    node[below,font=\sffamily\tiny,text=MetisGray]
    {\figtext{paired history edge}{配对历史边}} (assistant.south east);

  \begin{scope}[on background layer]
    \node[draw=MetisBlue,dashed,rounded corners=4pt,fit=(wire)(normal)(assistant)(telemetry),
      inner sep=4pt,label={[font=\sffamily\tiny,text=MetisBlue]above:\figtext{observation boundary}{观测边界}}] {};
    \node[draw=MetisAmber,dashed,rounded corners=4pt,fit=(permit)(exec)(result)(control)(lifecycle),
      inner sep=4pt,label={[font=\sffamily\tiny,text=MetisAmber]above:\figtext{effect boundary}{副作用边界}}] {};
  \end{scope}
\end{tikzpicture}%
}

\newcommand{\AgentLoopFigure}{%
\begin{tikzpicture}[node distance=4.5mm,font=\sffamily\tiny]
  \node[metisbox,fill=MetisPanel,text width=25mm] (compact)
    {\figtext{Prune / snip\\[-1pt]/ compact}{裁剪／截短／压缩}};
  \node[metisbox,fill=MetisSky,text width=25mm,below=of compact] (request)
    {\figtext{Build provider\\[-1pt]request}{构造 Provider 请求}};
  \node[metisbox,fill=MetisSky,text width=25mm,below=of request] (stream)
    {\figtext{Consume normalized\\[-1pt]stream}{消费归一化流}};
  \node[metisbox,fill=MetisMint,text width=25mm,below=of stream] (append)
    {\figtext{Persist assistant\\[-1pt]blocks}{持久化 Assistant 内容块}};
  \node[diamond,draw=MetisRule,fill=MetisSand,aspect=2.2,align=center,
        inner sep=2pt,below=5mm of append] (calls)
    {\figtext{tool uses?}{有工具调用？}};
  \node[metisbox,fill=MetisSand,text width=25mm,below left=7mm and 7mm of calls] (dispatch)
    {\figtext{Preflight and\\[-1pt]dispatch}{预检并调度}};
  \node[metisbox,fill=MetisMint,text width=25mm,below=of dispatch] (results)
    {\figtext{Append ordered\\[-1pt]results}{追加有序结果}};
  \node[metisbox,fill=MetisRose,text width=25mm,below right=7mm and 7mm of calls] (done)
    {\figtext{Emit terminal\\[-1pt]event}{发出终止事件}};

  \draw[metisarrow] (compact) -- (request);
  \draw[metisarrow] (request) -- (stream);
  \draw[metisarrow] (stream) -- (append);
  \draw[metisarrow] (append) -- (calls);
  \draw[metisarrow] (calls) -- node[above left]{\figtext{yes}{是}} (dispatch);
  \draw[metisarrow] (dispatch) -- (results);
  \draw[metisarrow] (results.west) -- ++(-5mm,0) |- (compact.west);
  \draw[metisarrow] (calls) -- node[above right]{\figtext{no}{否}} (done);
  \draw[metisdash] (done.east) -- ++(4mm,0) node[right,align=left]
    {\figtext{repair orphans on every return}{任一返回路径均修复孤儿调用}};
\end{tikzpicture}%
}

\newcommand{\PermissionFlowFigure}{%
\begin{tikzpicture}[node distance=5mm and 6mm]
  \node[metisbox,fill=MetisSky,text width=22mm,minimum height=11mm] (call)
    {\figtext{Invocation}{调用}\\[-1pt]\scriptsize tool + input\\[-1pt]+ path};
  \node[metisbox,fill=MetisRose,text width=24mm,minimum height=11mm,right=of call] (plan)
    {\figtext{Plan boundary}{Plan 边界}\\[-1pt]\scriptsize read/meta or deny};
  \node[metisbox,fill=MetisRose,text width=28mm,minimum height=11mm,right=of plan] (immune)
    {\figtext{Bypass-immune checks}{绕过免疫检查}\\[-1pt]\scriptsize safety paths\\[-1pt]+ secret reads};
  \node[metisbox,fill=MetisSand,text width=25mm,minimum height=11mm,right=of immune] (rules)
    {\figtext{Rule resolution}{规则解析}\\[-1pt]\scriptsize authority\\[-1pt]then recency};
  \node[metisbox,fill=MetisSand,text width=24mm,minimum height=11mm,right=of rules] (scope)
    {\figtext{Path scope}{路径作用域}\\[-1pt]\scriptsize cwd + add-dir};
  \node[metisbox,fill=MetisMint,text width=25mm,minimum height=11mm,right=of scope] (mode)
    {\figtext{Mode fallback}{模式回退}\\[-1pt]\scriptsize classifier\\[-1pt]if bypass};

  \draw[metisarrow] (call) -- (plan);
  \draw[metisarrow] (plan) -- (immune);
  \draw[metisarrow] (immune) -- (rules);
  \draw[metisarrow] (rules) -- (scope);
  \draw[metisarrow] (scope) -- (mode);

  \node[metisbox,fill=MetisMint,text width=31mm,below=12mm of rules] (allow)
    {\textbf{ALLOW}\\[-1pt]\figtext{admitted after batch preflight}{批量预检后准入}};
  \node[metisbox,fill=MetisSand,text width=31mm,left=8mm of allow] (ask)
    {\textbf{ASK}\\[-1pt]\figtext{resolve all prompts\\[-1pt]before effects}{所有询问先于副作用完成}};
  \node[metisbox,fill=MetisRose,text width=31mm,right=8mm of allow] (deny)
    {\textbf{DENY}\\[-1pt]\figtext{typed error result}{类型化错误结果}};
  \coordinate (outcome-bus-left) at ($(plan.south)+(0,-4mm)$);
  \coordinate (outcome-bus-right) at ($(mode.south)+(0,-4mm)$);
  \draw[line width=0.5pt,color=MetisGray]
    (outcome-bus-left) -- (outcome-bus-right);
  \foreach \a in {plan,immune,rules,scope,mode}
    \draw[dashed,line width=0.5pt,color=MetisGray]
      (\a.south) -- (\a.south |- outcome-bus-left);
  \foreach \a in {ask,allow,deny}
    \draw[metisdash] (\a.north |- outcome-bus-left) -- (\a.north);
\end{tikzpicture}%
}

\newcommand{\ContextLifecycleFigure}{%
\begin{tikzpicture}[node distance=5mm and 5mm]
  \node[metisbox,fill=MetisSky,text width=21mm,minimum height=12mm] (image)
    {\textbf{0}\\\figtext{Image prune}{图像裁剪}\\[-1pt]\scriptsize keep recent};
  \node[metisbox,fill=MetisSky,text width=22mm,minimum height=12mm,right=of image] (snip)
    {\textbf{1}\\\figtext{Snip}{截短}\\[-1pt]\scriptsize cheap, lossy};
  \node[metisbox,fill=MetisMint,text width=24mm,minimum height=12mm,right=of snip] (spill)
    {\textbf{2}\\\figtext{Disk offload}{磁盘卸载}\\[-1pt]\scriptsize recoverable\\[-1pt]pointer};
  \node[metisbox,fill=MetisSand,text width=24mm,minimum height=12mm,right=of spill] (collapse)
    {\textbf{3}\\\figtext{Middle collapse}{中段折叠}\\[-1pt]\scriptsize bounded summary};
  \node[metisbox,fill=MetisAmber!18,text width=24mm,minimum height=12mm,right=of collapse] (full)
    {\textbf{4}\\\figtext{Full compact}{完整压缩}\\[-1pt]\scriptsize boundary + tail};
  \node[metisbox,fill=MetisRose,text width=24mm,minimum height=12mm,right=of full] (guard)
    {\textbf{5}\\\figtext{Post guard}{压缩后保护}\\[-1pt]\scriptsize cap + retry};
  \foreach \a/\b in {image/snip,snip/spill,spill/collapse,collapse/full,full/guard}
    \draw[metisarrow] (\a) -- (\b);
  \draw[decorate,decoration={brace,amplitude=3pt},MetisGray]
    ($(image.south west)+(0,-6mm)$) -- ($(spill.south east)+(0,-6mm)$)
    node[midway,below=4pt,font=\sffamily\scriptsize]{\figtext{payload control}{载荷控制}};
  \draw[decorate,decoration={brace,amplitude=3pt},MetisGray]
    ($(collapse.south west)+(0,-6mm)$) -- ($(guard.south east)+(0,-6mm)$)
    node[midway,below=4pt,font=\sffamily\scriptsize]{\figtext{history restructuring}{历史重构}};
\end{tikzpicture}%
}

\hypersetup{
  pdftitle={Metis: Typed Runtime Mediation for Tool-Using Software Agents},
  pdfauthor={Jun Yu},
  pdfsubject={Public preprint},
  pdfkeywords={software engineering agents, tool-use runtime, permissions, concurrency, execution traces, context repair}
}

\begin{document}
\hypersetup{pdfauthor={Jun Yu}}

\title{Metis: Typed Runtime Mediation for Tool-Using Software Agents}

\author{Jun Yu\\
\small Independent Researcher, China\\
\small\texttt{ricardoporsche001@icloud.com}}
\date{}
\maketitle
\thispagestyle{plain}
\pagestyle{plain}

\begin{abstract}
Software agents connect probabilistic model output to operations that change repositories, processes, networks, and graphical applications. We present Metis, a multi-provider runtime that converts provider streams into typed events before admitted calls reach external effects. Its execution path makes permission decisions, interference classes, terminal results, and lifecycle transitions explicit and inspectable. We evaluate these mechanisms on frozen source artifacts. Across 30 matched real-I/O pairs, four-class mediation reduced median elapsed time from 25.958\,ms under forced serialization to 14.146\,ms. The mean paired difference was $-12.295$\,ms (95\% bootstrap interval $[-12.968,-11.694]$), with mediation faster in all pairs. A ten-case fault matrix exposed duplicate-identifier and rollback limits. In a child-boundary ablation, the full gate-plus-registry condition blocked the declared unauthorized effect and hid all five escape tools. Removing both protections reversed both observations. A decision-only permission oracle matched all ten declared cases across five invocation routes. Five model conditions also completed a fixed Read-marker protocol in 3/3 trials each. These results support bounded claims about dispatch, permission routing, child authority, and provider-valid trace closure. They do not establish model competence, semantic safety, rollback, or superiority over another runtime.
\end{abstract}


\section{Introduction}
\label{sec:introduction}

Software-engineering agents no longer only suggest code. They can read and modify repositories, invoke build and test processes, call remote services, and control graphical applications. This transition makes the runtime between a model and its environment a software-engineering object in its own right. A generated token can usually be ignored, while an admitted command or pointer action may already have changed external state.

Existing research primarily improves either the policy that proposes an action or the harness that exposes a task environment. ReAct interleaves language reasoning with environmental observations~\cite{yao2023react}; Toolformer learns where API calls improve future-token prediction~\cite{schick2023toolformer}; and coding-agent systems expose repositories and shells through specialized interfaces~\cite{yang2024sweagent,wang2024openhands}. These approaches make tool use and repository interaction possible, but they do not by themselves answer a downstream systems question: after a call is proposed, which component admits it, orders it relative to other calls, records its terminal result, and preserves a provider-valid history after interruption or context reduction?

A runtime at this boundary must resolve several coupled requirements. It must normalize incompatible provider streams into ordered calls, decide permission before effects across different invocation routes, and serialize interfering calls without suppressing safe concurrency. It must also pair every accepted tool-use identifier with a terminal result on supported error, panic, cancellation, and restart paths. Finally, it must reduce long histories and narrow child-agent authority without claiming process isolation or semantic preservation. Solving any one requirement in isolation leaves gaps: a valid provider request need not be authorized, and an authorized call need not produce an ordered, closed history.

\begin{figure}[t]
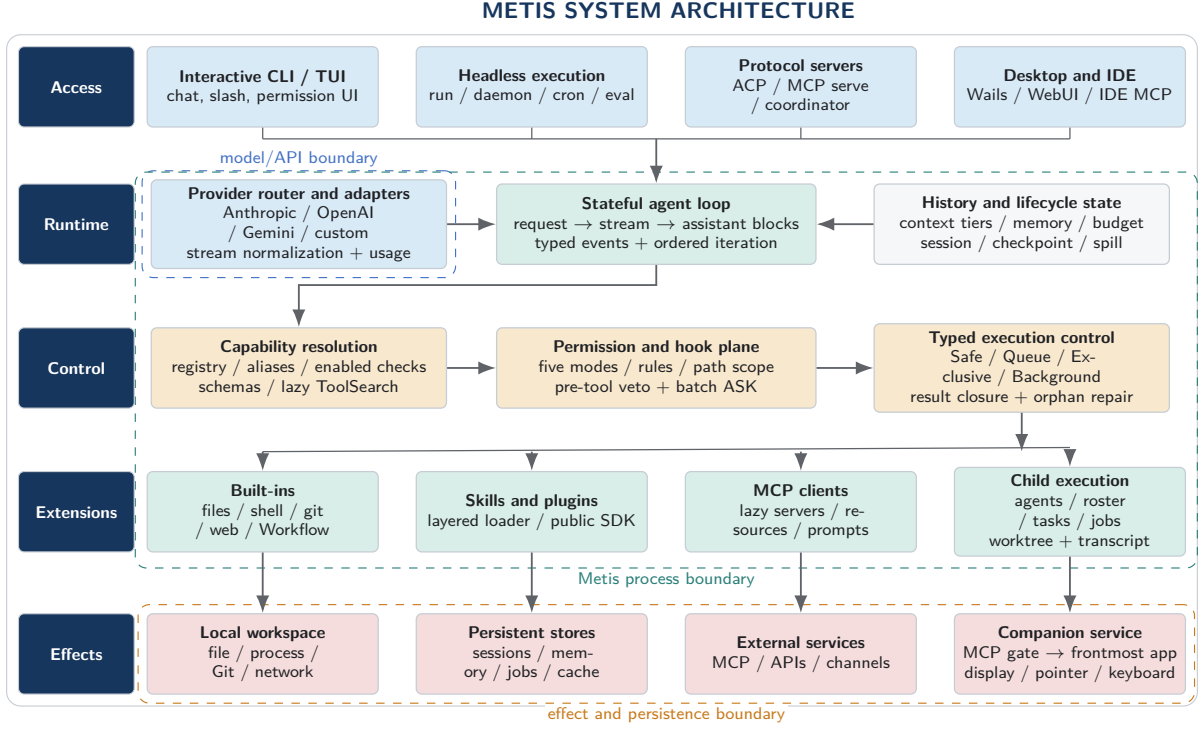

\centering
\resizebox{0.99\textwidth}{!}{\MetisArchitectureFigure}
\caption{\textbf{Metis system architecture.} Interactive, headless, protocol, and desktop entry points converge on a stateful loop. Provider adapters cross the model/API boundary. The permission and hook plane, capability resolver, and four-class dispatcher mediate registered effects. A separate companion implementation adds a frontmost-application gate for computer input.}
\label{fig:architecture}
\Description{A layered architecture diagram. Access surfaces feed provider adapters and a stateful agent loop. A registry, permission gate, and typed dispatcher mediate extensions before local, persistent, external, and computer-use effects. Dashed boxes mark model, process, and effect boundaries.}
\end{figure}

Metis addresses this gap by converting model-proposed calls into typed runtime events and a derived execution trace. Permission decisions, scheduling classes, terminal results, and lifecycle transitions become explicit edges in that trace. The trace is an audit object over runtime events, not a claim that model reasoning is verified. Figure~\ref{fig:architecture} expands this object into the full system: access surfaces, provider adaptation, lifecycle control, execution mediation, extensions, and external effects.

The separation is operational rather than cosmetic. Interactive approval, protocol permission replies, and headless policy handling terminate at the same gate. Built-ins, plugin tools, and MCP tools resolve through one registry before dispatch. Textual skills change instructions and tool visibility but do not create a second authority path. A child loop receives a cloned gate and filtered registry. The sequential \texttt{Workflow} tool gates every shell step through the Bash permission path; it is not a general dependency graph. Finally, the companion computer-use implementation is outside the Metis process and therefore adds, rather than replaces, the main runtime gate.

This paper makes four bounded contributions:
\begin{enumerate}
  \item It formulates runtime mediation for tool-using software agents as a typed event graph with explicit permission, scheduling, result, and lifecycle edges.
  \item It specifies and traces a five-mode permission order and a four-class dispatcher, including batch preflight before effects.
  \item It separates terminal-result closure from context retention and states the weaker identifier-coverage guarantee provided by orphan repair.
  \item It evaluates the mechanisms on frozen artifacts using a paired real-I/O ablation, fault matrix, child-boundary ablation, route-level permission oracle, and five model conditions with three trials each.
\end{enumerate}

The evaluation is deliberately mechanism-centered. Frozen source tests support named contracts, while controlled studies isolate real-I/O dispatch, fault handling, and child authority. Live model trials test only a minimal wire/runtime protocol. Four historical defects also have frozen fail-to-pass oracles. An exploratory maintenance pair is reported separately because it verifies treatment activation but cannot estimate an average effect. Repository size, commit activity, and tool counts are excluded because they do not validate runtime mechanisms. The evidence therefore supports local implementation and mechanism claims, not product-level safety or comparative maintenance effectiveness.

The remainder of the paper positions runtime mediation against adjacent agent and harness research, formalizes the threat model and event graph, describes the runtime mechanisms, evaluates their bounded claims, and closes with validity limits and the studies still required for broader conclusions.

\section{Related Work}
\label{sec:related}

The development of tool-using agents can be read as changes in the object that receives control: a textual action trajectory, a learned invocation policy, a repository harness, a delegated agent graph, and an effect-mediating runtime. These strata overlap rather than replace one another. Table~\ref{tab:related} compares where decisions and failures reside; it does not combine benchmark scores obtained with different models, tasks, or environments.

\subsection{Generated actions and repository harnesses}

ReAct made the reasoning--action--observation trajectory explicit: language thoughts update working context, while environment actions obtain new observations~\cite{yao2023react}. Toolformer moved part of tool choice into training by retaining linearized API calls that improve future-token prediction~\cite{schick2023toolformer}. Both works concern how a model proposes or exploits an action. They leave admission, interference, and terminal-result handling to the surrounding environment.

Repository agents such as SWE-agent and OpenHands made that environment a first-class software harness with files, commands, and feedback~\cite{yang2024sweagent,wang2024openhands}. The agentic-SDLC literature describes a broader shift from suggestion-oriented assistance to delegated repository work under supervision~\cite{bhati2026agenticsdlc}. Agent-adaptation research further separates changes to the model policy from changes to tools and feedback channels~\cite{jiang2025adaptation}. These distinctions motivate treating the runtime and harness as experimental objects independent of the model. Metis occupies this systems layer, but narrows its claim to mediation after a call has been proposed.

\subsection{Delegation and recursive harnesses}

Within delegation research, Recursive Agent Harnesses sharpen the distinction between a model policy and its execution environment by treating a complete tool-bearing harness, rather than a bare model call, as the recursive unit and holding the backbone fixed when estimating harness effects~\cite{lumer2026recursive}. A Metis child is likewise a complete loop, but with a cloned permission gate, filtered registry, and bounded capacity. The evidence here traces authority narrowing; it does not establish that recursion improves task success.

\subsection{Traces, retention, and capability representation}

The orchestration-trace literature represents multi-agent execution as a variable-shape event graph whose nodes include spawning, messaging, tool use, return, and aggregation~\cite{zhang2026orchestration}. That graph is an optimization and credit-assignment object. Metis uses a related mathematical form for a different purpose: permission, scheduling, result, and lifecycle nodes make a concrete runtime execution inspectable. No reward assignment or learned orchestrator is introduced.

Long-running systems separate protocol continuity from memory quality. MEMTIER studies episodic and semantic retention together with retrieval bottlenecks~\cite{sidik2026memtier}, while Neural Procedural Memory targets whether an agent enacts a procedure rather than merely retrieves text~\cite{zhao2026npm}. Metis currently provides payload control, spill files, lossy summaries, and archival retrieval. These mechanisms can preserve a structurally acceptable continuation without demonstrating factual recall or procedural transfer.

Tool representation is another layer. TSCG compiles tool schemas into a deterministic text representation~\cite{sakizli2026tscg}; Parametric Skills turns textual skills into test-time model adapters~\cite{zhao2026parametric}. Metis instead uses lazy schemas and source-ranked \texttt{SKILL.md} documents. This comparison identifies a deployment choice, not equivalence to schema compilation or model-internal skill acquisition.

\subsection{Permission coverage and adversarial effects}

An action-level study of Claude Code's permission gate shows why task completion is an inadequate safety unit: a task may succeed while individual state-changing actions cross an intended authorization boundary~\cite{ji2026permission}. It also exposes a coverage problem. Equivalent effects can be routed through tools that traverse different checks, so a gate must report which effect paths it actually mediates.

Prompt-centered and tool-output attacks probe complementary boundaries. Goal reframing can induce harmful actions despite fixed rule-following instructions~\cite{mouzouni2026exploitation}. AgentRedBench places adversarial content in read integrations and observes whether it propagates into downstream write actions~\cite{dingeto2026agentredbench}. Metis places path, tool, and frontmost-application checks outside the prompt, but matched versions of these evaluations have not been run. The present claim is architectural coverage with explicit residuals, not a transferred defense rate.

\subsection{Systems mediation, workflows, and recovery}

Runtime mediation predates LLM agents. Classic protection principles identify complete mediation and least privilege as design criteria~\cite{saltzer1975protection}. Workflow research catalogs recurring sequencing, synchronization, and cancellation structures~\cite{vanderAalst2003workflow}, while long-running transaction research uses explicit compensation for partial effects~\cite{garciamolina1987sagas}. These traditions bound the present contribution. Metis integrates permission coverage, interference classes, and provider-valid terminal results for model-proposed calls; it does not claim to invent authorization, general workflow control, or transactional compensation.

\begin{table}[H]
\caption{\textbf{Locus of control in tool-using agent systems.} Rows are overlapping research lineages, not a leaderboard. ``Effect boundary'' names where a proposed action first meets an execution environment.}
\label{tab:related}
\Description{A four-column comparison of reason-act prompting, learned API use, repository harnesses, recursive harnesses, and Metis runtime mediation. It contrasts their control object, effect boundary, retained state, and relation to Metis.}
\centering
\small
\renewcommand{\arraystretch}{1.05}
\setlength{\tabcolsep}{3.5pt}
\begin{tabularx}{\textwidth}{@{}B{27mm}L{36mm}Y L{37mm}@{}}
\toprule
Lineage & Control object and decision & Effect boundary and retained state & Position relative to Metis \\
\midrule
Reason--act prompting~\cite{yao2023react} & In-context model policy over a thought--action--observation trajectory & Task environment interprets actions; prompt retains observations & Upstream of admission and execution \\
Learned API use~\cite{schick2023toolformer} & Trained sequence policy selects embedded calls & API wrapper returns text to model context & Upstream invocation policy \\
Repository harness~\cite{yang2024sweagent,wang2024openhands} & Model acts through an issue-solving repository interface & Harness supplies shell, files, sandbox, trajectory, and workspace & Neighboring layer; comparison requires fixed model and task \\
Recursive harness~\cite{lumer2026recursive} & Parent harness delegates complete subproblems & Each child retains tools; summaries return through recursive context & Metis narrows a child harness; benefit remains unevaluated \\
Runtime mediation (Metis) & Model proposes; gate admits; dispatcher orders a typed event graph & Permission and scheduling precede effects; session, spill, repair, and child state persist & Runtime object evaluated in this paper \\
\bottomrule
\end{tabularx}
\end{table}

\section{Problem Formulation and Threat Model}

Let provider $p$ produce a stream $S_p$. A normalizer maps that stream to an ordered sequence of shared content blocks,
\begin{equation}
  \mathcal{N}_p(S_p)=B=(b_1,\ldots,b_m).
  \label{eq:normalize}
\end{equation}
A tool-use block is $u=(q,n,x)$, where $q$ is its identifier, $n$ its tool name, and $x$ its structured input. The runtime must decide whether $u$ may cause an effect, schedule it, and return a paired result.

One run induces a typed directed graph
\begin{equation}
  \mathcal{G}=(V,E,\lambda_V,\lambda_E),
  \label{eq:graph}
\end{equation}
where vertices are runtime events and edges encode order, decision, pairing, or lifecycle dependence. This graph is reconstructed from typed events and identifiers. Metis does not persist a graph database for every run.

The runtime aims to maintain four local properties. \emph{Authorization-before-effect} requires a final permission decision before execution. \emph{Ordered interference} requires calls with shared mutation hazards to respect their class. \emph{Terminal-result closure} requires a result for each accepted tool-use identifier within the supported failure model. \emph{Bounded continuation} requires the next provider request to remain structurally valid after context control or repair. Table~\ref{tab:trust} states the assumptions and non-guarantees at each trust boundary.

\begin{table}[H]
\caption{\textbf{Trust boundaries and non-guarantees.} Each row separates a runtime assumption from a property that the present evidence does not establish.}
\label{tab:trust}
\Description{A three-column table mapping six boundaries to runtime assumptions and properties not established by this paper.}
\centering
\small
\renewcommand{\arraystretch}{1.05}
\begin{tabularx}{\columnwidth}{@{}B{24mm}L{42mm}Y@{}}
\toprule
Boundary & Runtime assumption & Not established \\
\midrule
Model & Emits parseable content or tool blocks & Correct reasoning or intent \\
Provider & Adapter observes available stream fields & Identical semantics across providers \\
Policy & Metadata, rules, and scope are configured correctly & Complete semantic authorization \\
Tool & Declares permission and concurrency needs correctly & Absence of hidden side effects \\
Host & Process and required storage remain available & Recovery after host loss \\
GUI & Frontmost-app query reflects the target & Intent, focus stability, or visual correctness \\
\bottomrule
\end{tabularx}
\end{table}

The threat model includes malformed or adversarial model-proposed calls, misleading tool output, misconfigured policy, incorrect concurrency metadata, tool error or panic, cancellation, provider interruption, context pressure, and child propagation. The trusted computing base includes the gate and dispatcher implementation, declared tool metadata, required persistence, provider fields exposed to adapters, and operating-system queries used by computer input. The paper does not claim protection from a compromised host, a malicious tool that lies about its side effects, semantic interpretation of user intent, transactional rollback, or arbitrary process/host loss.

\section{Runtime Design}

\subsection{Typed events and provider normalization}

\begin{figure}[t]
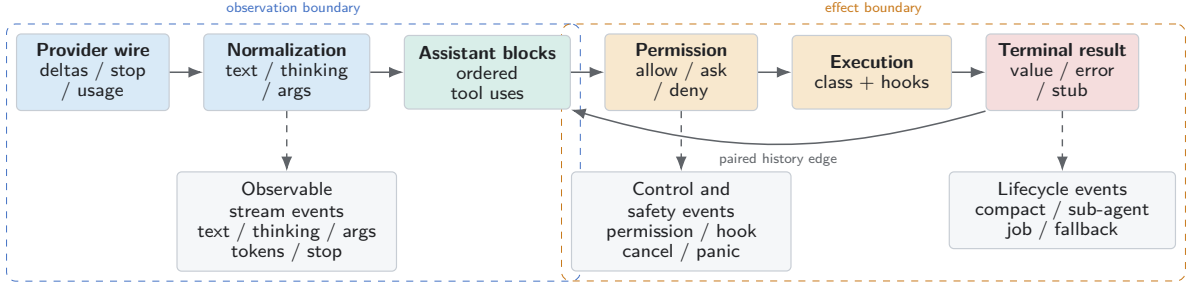

\centering
\resizebox{0.98\textwidth}{!}{\RuntimeEventGraphFigure}
\caption{\textbf{Runtime event graph.} Provider deltas become ordered assistant blocks. Permission and execution events cross the effect boundary. A paired result returns to history, while compaction, fallback, and child-loop events expose lifecycle changes.}
\label{fig:eventgraph}
\Description{A left-to-right event graph from provider wire events through normalization, assistant blocks, permission, execution, and terminal result. Side boxes group telemetry, control, and lifecycle events.}
\end{figure}

Metis emits events for model output, tool progress, permissions, lifecycle transitions, and exceptional conditions. Figure~\ref{fig:eventgraph} separates observation from effect regions. Provider normalization preserves content order while resolving incremental details. A provider may emit complete tool calls, incremental JSON arguments, or separate thinking blocks. The adapter accumulates these forms into shared blocks before dispatch. The invariant is ordering and parseable shared structure for implemented adapter cases, not semantic identity between provider APIs.

For adjacent normalized blocks, define $b_i\prec_B b_j$ when the adapter observed completion of $b_i$ before $b_j$. The request builder preserves this order:
\begin{equation}
  b_i\prec_B b_j \Longrightarrow
  \operatorname{pos}_B(b_i)<\operatorname{pos}_B(b_j).
  \label{eq:blockorder}
\end{equation}
Equation~\ref{eq:blockorder} states an ordering invariant, not semantic equivalence. Targeted stream tests cover interleaved parallel tools, thinking flushes, and lost-byte resynchronization. They validate implemented adapter cases, rather than all provider behaviors.

\subsection{Loop state machine}

\begin{figure}[t]
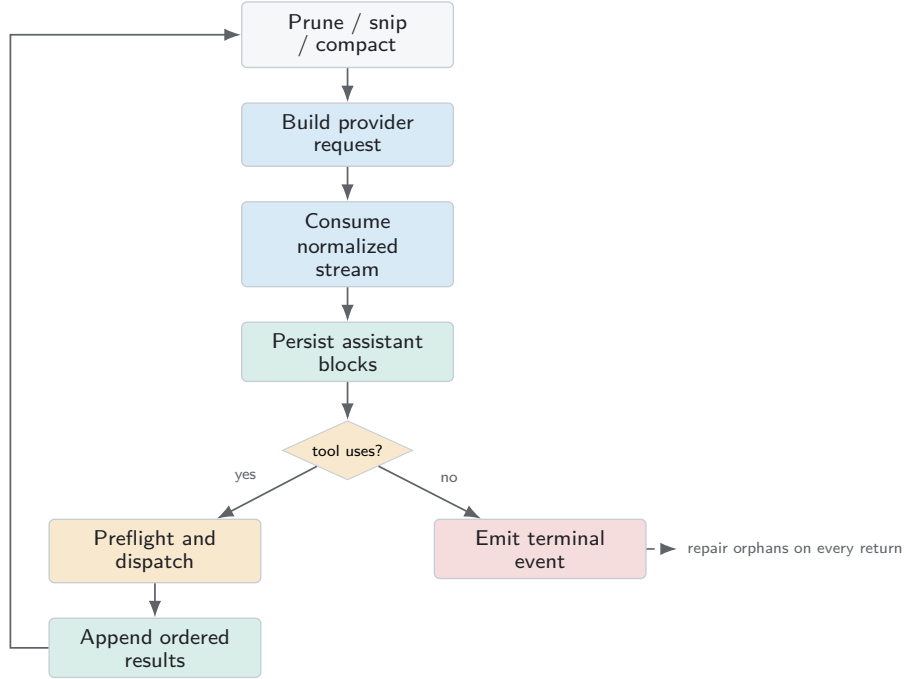

\centering
\resizebox{0.75\columnwidth}{!}{\AgentLoopFigure}
\caption{\textbf{Agent-loop state machine.} Context control precedes request construction. Normalized assistant blocks are persisted before preflight. Return paths invoke orphan repair as defense in depth.}
\label{fig:agentloop}
\Description{A flowchart from context control to request, stream, assistant persistence, a tool-use decision, either dispatch and ordered results looping back, or a terminal event.}
\end{figure}

The loop alternates context preparation, provider streaming, assistant persistence, and tool dispatch. A turn without tool calls emits a terminal event. A turn with calls appends ordered results and begins another iteration. Figure~\ref{fig:agentloop} summarizes the control flow and its repair path.

Let $H_t$ be persisted history and $B_t$ the assistant blocks from iteration $t$. The abstract transition is
\begin{equation}
  H_{t+1}=
  \begin{cases}
    C(H_t)\oplus B_t, & U(B_t)=\varnothing,\\
    C(H_t)\oplus B_t\oplus D(B_t), & \text{otherwise},
  \end{cases}
  \label{eq:loop}
\end{equation}
where $C$ is the context controller and $D$ the permissioned dispatcher. The operator $\oplus$ denotes ordered message extension. Equation~\ref{eq:loop} is an implementation abstraction, not a claim that compaction preserves every semantic detail.

\subsection{Permission semantics and batch preflight}

\begin{figure}[t]
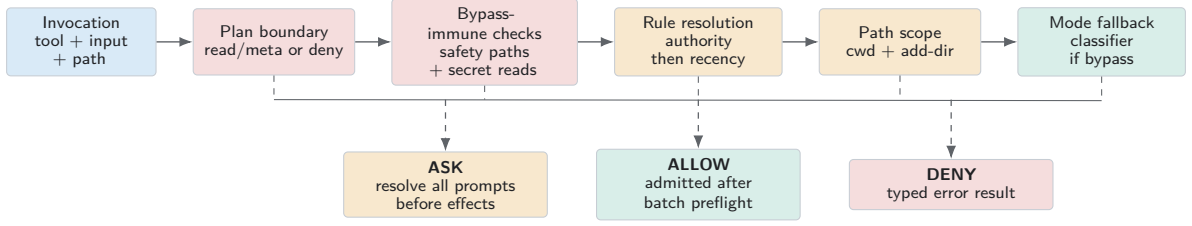

\centering
\resizebox{0.98\textwidth}{!}{\PermissionFlowFigure}
\caption{\textbf{Permission resolution and batch preflight.} The gate evaluates the plan boundary, bypass-immune checks, rule precedence, path scope, and mode fallback in order. The first resolving check feeds a shared ALLOW/ASK/DENY outcome bus. Every ASK in a batch resolves before an admitted effect starts. A denial breaker can degrade repeated automated DENY decisions to ASK; classifier failure in explicit bypass mode remains a documented fail-open boundary.}
\label{fig:permissionflow}
\Description{A flowchart from invocation through ordered plan, bypass-immune, rule, scope, and mode checks. The first resolving check feeds a common bus that branches to allow, ask, or deny.}
\end{figure}

For call $u$ under state $\sigma$, the gate returns
\begin{equation}
  d(u,\sigma)\in\{\mathsf{allow},\mathsf{ask},\mathsf{deny}\}.
  \label{eq:permission}
\end{equation}
The state includes session mode, matching rules, path scope, and input-sensitive checks. The five public modes are \texttt{default}, \texttt{acceptEdits}, \texttt{bypassPermissions}, \texttt{plan}, and \texttt{dontAsk}. Plan mode is evaluated before ordinary rules, so a stale allow cannot authorize mutation. Bypass-immune path and secret-read checks follow for eligible calls. Rules resolve by authority and recency. Path scope is checked before the mode fallback. In bypass mode, dangerous command patterns precede the optional classifier. Figure~\ref{fig:permissionflow} shows this ordered resolution and the batch-level outcomes.

The precedence can be written as
\begin{equation}
 d(u,\sigma)=
 \begin{cases}
 d_{\mathrm{plan}}, & m=\mathsf{plan},\\
 d_{\mathrm{immune}}, & I_{\mathrm{immune}}(u)=1,\\
 d_{\mathrm{rule}}, & \exists r\ \mathrm{match}(r,u),\\
 d_{\mathrm{scope}}, & \mathrm{outside}(u,\sigma)=1,\\
 d_{\mathrm{mode}}, & \mathrm{otherwise}.
 \end{cases}
 \label{eq:precedence}
\end{equation}
Each branch may contain its own guard. Equation~\ref{eq:precedence} specifies precedence, not a stateless classifier.

Let $A_B$ be calls in batch $B$ whose initial decision is \textsf{ask}, and $P_B$ calls eventually admitted. Metis resolves every pending question before an admitted effect begins:
\begin{equation}
  \max_{u\in A_B}\tau_{\mathrm{decision}}(u)
  \leq
  \min_{v\in P_B}\tau_{\mathrm{start}}(v).
  \label{eq:preflight}
\end{equation}
Equation~\ref{eq:preflight} requires all pending batch decisions to precede the first admitted effect. A rejected call receives a typed error result, so denial does not create a hole in the next request. Bypass is not equivalent to removing the gate: hard checks remain, but a classifier error fails open after them because the user explicitly selected bypass. This is a documented boundary, not a general fail-closed guarantee.

\subsection{Four-class scheduling}

\begin{figure}[t]
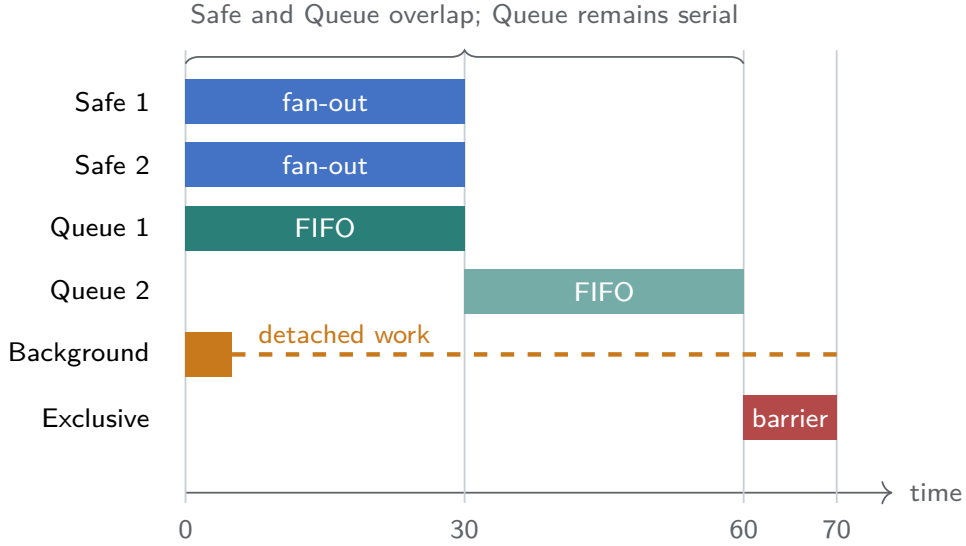

\centering
\resizebox{0.82\columnwidth}{!}{\DispatchTimelineFigure}
\caption{\textbf{Four-class scheduling semantics.} Queue remains FIFO while overlapping Safe calls. Exclusive calls form a barrier. Background calls return a handshake before detached work completes.}
\label{fig:timeline}
\Description{A timeline with two parallel Safe calls, two serial Queue calls that overlap Safe work, a detached Background handshake, and an Exclusive call after a barrier.}
\end{figure}

Every admitted call receives an input-sensitive class
\begin{equation}
  \kappa(u)\in\{\mathsf{Safe},\mathsf{Queue},\mathsf{Exclusive},\mathsf{Background}\}.
  \label{eq:classes}
\end{equation}
The same tool may classify different inputs differently. Safe calls fan out. Queue calls use one FIFO worker while overlapping Safe calls. Exclusive calls wait for the earlier wave and run serially. Background calls return a handshake without joining detached completion to the foreground path. Figure~\ref{fig:timeline} visualizes the intended overlap and barrier semantics.

For service times $t_i$, handshake times $h_i$, and foreground sets $S,Q,X$, the idealized critical path is
\begin{equation}
  T^*=\max\left(\max_{i\in S}t_i,\sum_{i\in Q}t_i,\sum_{i\in B_g}h_i\right)+\sum_{i\in X}t_i.
  \label{eq:critical}
\end{equation}
In Equation~\ref{eq:critical}, $B_g$ contains Background handshakes, not detached job durations. The formula assumes sufficient resources and no undeclared dependency between Safe and Queue calls. The empirical study below does not estimate this idealized quantity from programmed sleeps; it directly compares paired wall-clock observations from declared scheduling classes and a forced-serial ablation under the same real-I/O workload.

\subsection{Terminal-result closure and repair}

Dispatch converts failure into data visible to the next model turn. Unknown tools, hook short-circuits, denials, rejected questions, pre-run cancellation, panic, ordinary error, invalid nil results, and success all produce a result block with the original identifier in supported in-process paths. Blocks return in input-call order even when Safe calls finish out of order.

Let $\mathbf{U}_B=(q_1,\ldots,q_k)$ list call identifiers in input order and let $\mathbf{R}_B=(\hat q_1,\ldots,\hat q_\ell)$ list returned-result identifiers. If the process survives and batch handling returns, the dispatcher targets
\begin{equation}
  \lvert\mathbf{R}_B\rvert=\lvert\mathbf{U}_B\rvert
  \quad\text{and}\quad
  \forall i\in\{1,\ldots,k\},\ \hat q_i=q_i.
  \label{eq:batchclosure}
\end{equation}
Equation~\ref{eq:batchclosure} is a per-call sequence property, so it preserves multiplicity and order even when identifiers are duplicated. It neither establishes identifier uniqueness nor rolls back an external side effect, and it does not survive arbitrary host termination.

\begin{figure}[t]
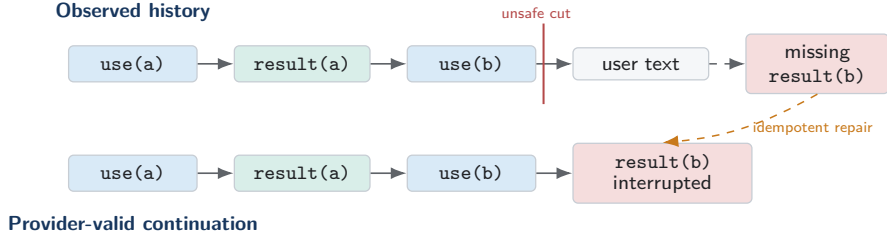

\centering
\resizebox{0.74\textwidth}{!}{\HistoryClosureFigure}
\caption{\textbf{History closure after interruption or compaction.} Boundary adjustment or an interrupted-result stub restores identifier coverage without reconstructing or rolling back an effect.}
\label{fig:repair}
\Description{Two histories compare an unsafe cut after a tool use with a repaired history that appends an interrupted result for the missing identifier.}
\end{figure}

Cancellation between assistant persistence and result persistence can leave an orphaned tool-use identifier. Metis appends an interrupted-result stub for each identifier with no observed result. Figure~\ref{fig:repair} contrasts an unsafe cut with the repaired continuation.

For history $H$, let $U(H)$ and $R(H)$ be the sets of tool-use and result identifiers. Repair $\mathcal{R}$ supports
\begin{equation}
  U(H)\subseteq R(\mathcal{R}(H)),\qquad
  \mathcal{R}(\mathcal{R}(H))=\mathcal{R}(H).
  \label{eq:repair}
\end{equation}
Equation~\ref{eq:repair} states identifier coverage and idempotence, weaker than chronological one-to-one matching. A global satisfied-identifier set means duplicate identifiers or stale earlier results can hide temporal ambiguity.

\subsection{Context lifecycle}

\begin{figure}[H]
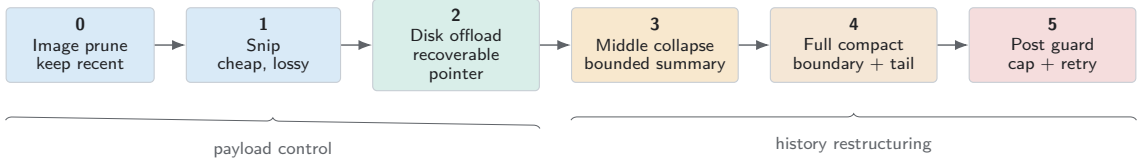

\centering
\resizebox{0.94\textwidth}{!}{\ContextLifecycleFigure}
\caption{\textbf{Context lifecycle.} Six stages escalate from local payload controls to LLM-backed history restructuring. They preserve protocol structure only; a post-compaction guard and overflow retry prevent one retained result from immediately exhausting the window again.}
\label{fig:context}
\Description{Six sequential stages show image pruning, snipping, disk offload, middle collapse, full compaction, and a post-compaction guard, grouped into payload control and history restructuring.}
\end{figure}

Context pressure triggers progressively stronger transformations: image pruning, result snipping, disk offload, bounded middle collapse, full compaction, and a post-compaction cap/retry. Let $q(H)$ estimate input tokens. A successful transformation $C_k$ is intended to satisfy
\begin{equation}
  q(C_k(H))\leq q(H),
  \label{eq:contextbudget}
\end{equation}
while preserving the active task anchor and valid use/result structure. Equation~\ref{eq:contextbudget} does not imply semantic equivalence. Snipping and summarization are intentionally lossy; spill recovery depends on the cached file remaining available. Boundary adjustment and orphan repair prioritize protocol acceptance, not the truth or completeness of a generated summary. Figure~\ref{fig:context} orders the six increasingly strong controls.

\subsection{Sub-agents and computer-use boundaries}

Metis can spawn a cold child loop or a fork that inherits a parent snapshot. Let $T_p$ be parent-visible tools, $T_r$ a profile allowlist, $T_c$ a call-site allowlist, and $D_r$ a profile denylist. The child surface is
\begin{equation}
  T_{\mathrm{child}}=(T_p\cap T_r\cap T_c)\setminus D_r,
  \label{eq:childtools}
\end{equation}
where a missing allowlist is the identity set. A child receives a cloned gate with fresh denial/memo state; rules and scope hooks are inherited. Optional worktree isolation separates repository writes, not network access, provider state, host resources, or credentials. Shared budget and spill state likewise do not imply process isolation.

The companion computer-use service classifies frontmost applications and input operations with
\begin{equation}
  \mathsf{Read}\prec\mathsf{Click}\prec\mathsf{Full},
  \label{eq:lattice}
\end{equation}
and admits operation $o$ for frontmost application $a$ when
\begin{equation}
  \operatorname{allow}(o,a)=\I[\operatorname{rank}(h(a))\geq\operatorname{rank}(r(o))].
  \label{eq:cuallow}
\end{equation}
Pointer actions require Click; typing, key combinations, clipboard writes, and application launch require Full. A frontmost lookup error fails closed for input. Unknown applications default to Click on the audited macOS path. The 250\,ms default cache creates a time-of-check to time-of-use focus window. Tests cover lattice logic and lookup failures, not visual grounding or input-delivery reliability.

\section{Evaluation}

\subsection{Research questions and evidence levels}

The evaluation distinguishes four questions:
\begin{samepage}
\begin{description}
  \item[RQ1: Snapshot conformance.] What do the frozen test logs establish about the runtime and companion implementation?
  \item[RQ2: Execution mechanisms.] What do paired real-I/O and injected-fault observations establish about scheduling and terminal closure?
  \item[RQ3: Authority boundaries.] How do child-boundary and route-level permission decisions change under controlled conditions?
  \item[RQ4: Model protocol and exploratory trace.] Do five model conditions complete one fixed tool protocol, and does one maintenance pair verify treatment activation?
\end{description}
\end{samepage}

These questions use different evidence levels. Frozen logs describe one source state, while controlled ablations isolate named runtime mechanisms. The decision-only oracle never executes its proposed effects, and the provider protocol exercises a trivial marker task. The exploratory maintenance trace contains one task, one repetition, and a strict whole-tree success rule. None of these studies establishes product-level reliability, semantic safety, model ability, or comparative maintenance effectiveness. Table~\ref{tab:evidence} maps every evaluated claim to its reproduction unit and evidence boundary.

\begin{table}[H]
\caption{\textbf{Claim-to-evidence map.} A pass supports only the named mechanism and boundary.}
\label{tab:evidence}
\Description{A three-column table maps eight runtime claims to frozen tests or controlled studies, reproduction units, and explicit evidence boundaries.}
\centering
\small
\renewcommand{\arraystretch}{1.05}
\setlength{\tabcolsep}{4pt}
\begin{tabularx}{\textwidth}{@{}B{31mm}L{55mm}Y@{}}
\toprule
Claim & Evidence and reproduction unit & Supported boundary \\
\midrule
Provider normalization & Frozen tests plus five model conditions, three marker trials each & Simple Read protocol; no semantic equivalence \\
Permission before effect & Frozen tests and ten decision-only cases over five routes & Handcrafted decisions; no executed effects \\
Four scheduling classes & 30 paired four-class/forced-serial real-I/O rows & One runtime and workload; no competing runtime \\
Terminal-result closure & Ten injected success/fault/restart cases & In-process structural closure; three observed negatives \\
Context lifecycle & Compaction, snip, overflow, and repair test paths & Structural continuation, not summary fidelity \\
Child narrowing & Two deterministic full-versus-ablated conditions & Combined gate/registry treatment; no host isolation \\
Computer-use lattice & Frozen tier and enforcement tests & Gate logic, not full GUI reliability \\
Maintenance provenance and pilot & Four frozen buggy/corrected pairs plus one evaluable baseline--mediated run pair & Task oracles plus one descriptive contrast; no population effect \\
\bottomrule
\end{tabularx}
\end{table}

\subsection{Frozen snapshots and test baseline}

All mechanism results refer to frozen snapshots of the main runtime, companion implementation, and permission route-closure variant. They are bound to a private evidence manifest; because public deposition remains future work, this preprint does not claim that the package is available.

The baseline counts terminal events in Go JSON logs. For Metis, leaf events were 4,998 pass, 2 fail, and 31 skip; counts including parent subtests were 5,178, 2, and 31. Its produced aggregate coverage profile was 63.5\%, with six observed packages absent, so it is not a whole-tree certificate. The companion produced 297/6/1 leaf and 306/6/1 named pass/fail/skip events. Its 64.2\% profile is explicitly partial because the platform package was interrupted; one named test started without a terminal event.

The two Metis failures had process-tree precondition signatures. Companion failures included clipboard/OCR environment signatures and stale count/MIME expectations. These classifications describe log symptoms, not proven root causes, and neither full suite is reported as passing.

\subsection{Paired real-I/O protocol and analysis}

Each matched pair ran the same five calls under two conditions: the declared Safe, Queue, and Exclusive classes, or a forced-serial variant that labeled every call Exclusive. The calls performed a filesystem read and hash, Git status, two loopback HTTP requests (each with a controlled 6\,ms server delay), and a filesystem write. Pair order alternated across 30 pairs. All 300 tool results completed without a tool error. The study ran on an Apple M2 Pro host with 12 CPU cores and 16\,GB of memory, macOS 26.5.2, and Go 1.26.1. The primary contrast is paired elapsed time, mediated minus forced serial. We report condition medians, the mean paired difference, a deterministic 20,000-resample percentile bootstrap interval over the 30 pairs (seed 20270813), and the pairwise direction count. This is a within-runtime mechanism ablation, not a comparison against another runtime or a maintenance speedup.

\subsection{RQ1: The frozen baseline is broad but not clean}

The baseline supplies a reproducible inventory and retains all failures. It supports source- and test-level tracing for the mechanisms in Table~\ref{tab:evidence}, but neither coverage percentage means complete repository coverage, parent and leaf events are not independent samples, and the interrupted companion run precludes a full-suite claim.

\subsection{RQ2: Real-I/O mediation reduced elapsed time in this workload}

Table~\ref{tab:dispatch} reports the paired real-I/O result. Four-class mediation had a 14.146\,ms median; forcing all calls serial had a 25.958\,ms median. The mean paired difference was $-12.295$\,ms, with a 95\% bootstrap interval of $[-12.968,-11.694]$\,ms. The mediated condition was faster in 30/30 pairs. The ratio of condition medians was 1.835, but we treat the paired difference as primary.

\begin{table}[H]
\caption{\textbf{Paired real-I/O dispatcher ablation.} The study contains $30$ matched pairs. Times are milliseconds; the difference is mediated minus forced serial.}
\label{tab:dispatch}
\Description{Two conditions list their medians, followed by the mean paired difference, its bootstrap interval, and the direction count.}
\centering
\small
\begin{tabular}{@{}lr@{}}
\toprule
Quantity & Estimate \\
\midrule
Four-class median & 14.146 \\
Forced-serial median & 25.958 \\
Mean paired difference & $-12.295$ \\
95\% paired-bootstrap interval & $[-12.968,-11.694]$ \\
Four-class faster & 30/30 pairs \\
\bottomrule
\end{tabular}
\end{table}

The 1.835 ratio is not reported as a general speedup: the forced-serial condition is an ablation within Metis, the loopback service includes a programmed delay, the task has deliberate overlap, and the study covers one host and workload.

The ten-case fault matrix complements timing by exercising success, returned error, nil result, panic, timeout, cancellation before Exclusive, partial mutation, and restart paths. \mbox{A normal orphan} was repaired after restart, and the Background path returned a paired handshake before its effect. Three negative results constrain the closure claim: duplicate IDs yielded two result blocks but only one unique terminal ID; a write followed by failure left residual state; and restart with a duplicate ID and one result did not achieve one-to-one terminal closure. Metis therefore provides neither duplicate-ID uniqueness nor transactional rollback.

\subsection{RQ3: Controlled authority evidence is narrow}

The child ablation compared two deterministic conditions. With both the child permission gate and plan-filtered registry, the declared-unauthorized effect was blocked and 0/5 escape tools were visible. Removing both protections admitted the effect and exposed 5/5. Because the treatment removes two protections together and has only one deterministic case per condition, it demonstrates a boundary consequence, not an average independent effect of either component.

The permission oracle contains one authorized and one unauthorized handcrafted state-change decision for each of five routes: built-in write, Bash, alias, MCP, and plugin-loaded MCP. It calls \texttt{CanUse} only; no state-changing \texttt{Execute} method runs. All ten decisions matched the oracle: five true positives and five true negatives, with no false positives or false negatives. This is a route-closure check over declared cases, not empirical calibration for arbitrary tools or user intent. Its permission route-closure variant is distinct from the exploratory trace's public-regression-feedback treatment.

\subsection{RQ4: Model protocol compatibility and an exploratory maintenance trace}

Five model conditions each ran three fresh sessions under the same marker prompt, Read schema, fixture, budget, and analyzer: \emph{gemini-3.5-flash}, \emph{ark-code-latest}, \emph{sensenova-6.8-flash-lite}, \emph{glm-5.2}, and \emph{deepseek-v4-flash}. A pass required exactly one Read call, a paired result with unique identifier closure, marker presence in the result, and marker return in final text. All five conditions passed 3/3, for 15/15 retained trials. An earlier \emph{gemini-3.5-flash} run was excluded before comparison because the analyzer read the wrong persisted result field; the corrected analyzer and unique sessions were used for its retained rerun. A sixth configured model, \emph{sensenova-u1-fast}, returned an availability error (HTTP 404: \texttt{model is not found}) on all three launch attempts before a protocol trace existed. We therefore treat those attempts as an availability exclusion, not a protocol outcome. The retained observations establish compatibility with one trivial wire/runtime protocol, not equivalence between models or model ability.

Four historical maintenance tasks are frozen separately. Each records a buggy revision, its direct corrected child, a prompt, allowed paths, a patch fixture, and a task-specific oracle. The tasks cover panic containment, duplicate task identifiers, a symlinked-home guard, and tool-error rendering. A fresh offline replay on 16 August 2026 observed the expected buggy failure and corrected-child pass for all four. This validates task construction only.

The maintenance treatment is a runtime intervention, not an extra model prompt. After each newly observed, non-empty mutation epoch within the declared workspace paths, the mediated condition runs a predeclared public regression command and injects bounded standard-output and standard-error diagnostics into the same model context. The baseline uses the same runtime with this hook disabled. The treatment receives no hidden oracle, gold patch, or fixture path; those artifacts are reserved for outcome scoring. Each mediated activation is recorded as an ordered \texttt{mutation\_observed}, \texttt{verifier\_started}, \texttt{verifier\_finished}, and \texttt{feedback\_injected} chain bound to the workspace state and public command hash.

One real-model pair then addressed panic containment with \emph{sensenova-6.8-flash-lite}. We fixed the model, runtime, adapter, prompt, tool schema, host, 900\,s timeout, 30 primary iterations plus up to two rescues, and 2,000,000/20,000 input/output-token caps. Only the public-regression-feedback treatment varied; a seeded permutation placed baseline first. \mbox{Both cells exited} the adapter normally, remained within budget and scope, had closed traces, and passed the isolation and publication firewalls. The mediated cell recorded three complete, workspace-bound treatment activations; the baseline recorded none.

Table~\ref{tab:maintenancepilot} reports the descriptive outcome. The baseline patch failed its hidden oracle and targeted regression, and it left \texttt{dispatch.go} uncompilable. The mediated patch passed both task-local checks. Its whole-tree run nevertheless failed four non-target tests in two packages, with configuration or host-I/O signatures. The preregistered whole-tree rule therefore classified both cells as unsuccessful. Relative to baseline, the mediated trajectory used 253.175\,s less wall time, 596,621 fewer input tokens, 6,192 fewer output tokens, and 14 fewer tool calls. These within-pair differences are observations, not effect estimates. One ordered pair supplies neither replication nor uncertainty and cannot separate treatment, model stochasticity, cache, or path effects.

\begin{table}[H]
\caption{\textbf{Exploratory single-pair maintenance trace.} Strict success required scope, the hidden oracle, the targeted regression, and a clean whole-tree run.}
\label{tab:maintenancepilot}
\Description{Baseline and mediated outcomes for one panic-containment run pair, including task-local and whole-tree success, time, tokens, tool calls, closure, and treatment activation.}
\centering
\small
\begin{tabularx}{\columnwidth}{@{}Xrr@{}}
\toprule
Observation & Baseline & Mediated \\
\midrule
Hidden oracle & Fail & Pass \\
Targeted package regression & Fail & Pass \\
Task-local success & 0/1 & 1/1 \\
Whole-tree regression & Fail & Fail \\
Whole-tree strict success & 0/1 & 0/1 \\
Wall time (s) & 556.622 & 303.447 \\
Input tokens & 884,363 & 287,742 \\
Output tokens & 15,301 & 9,109 \\
Tool calls & 33 & 19 \\
Trace closure & Pass & Pass \\
Complete treatment activations & 0 & 3 \\
\bottomrule
\end{tabularx}
\end{table}

Earlier quota-exhausted cells ended with exit code 4 and remain infrastructure exclusions. A paired counterpart that never launched is not imputed. Synthetic writer and analyzer checks also remain excluded. The new pair verifies that the treatment activated on three workspace mutations, but it does not close the maintenance-effectiveness question.

\section{Discussion}

\subsection{Policy, harness, and runtime are separate objects}

The related-work comparison suggests a three-part decomposition. A \emph{policy} proposes an action, a \emph{harness} supplies tools and feedback, and a \emph{runtime} decides how an admitted action crosses into effects. ReAct and Toolformer primarily modify the first object~\cite{yao2023react,schick2023toolformer}; repository and recursive agents expose the second~\cite{yang2024sweagent,lumer2026recursive}; Metis isolates mechanisms in the third. This decomposition prevents two inference errors. Improved task success under a new model does not validate runtime safety, and a stricter runtime does not establish better reasoning.

\subsection{The maintenance pair is activation evidence, not an effect estimate}

The exploratory pair shows that the frozen treatment loaded and activated on three workspace mutations. It also coincided with a task-local fail-to-pass contrast for one fixed model. Task-local correctness and whole-tree cleanliness disagreed because the mediated patch passed its task checks while four non-target tests failed. Reporting only the oracle would overstate success. Reporting only the strict binary would hide the observed repair and treatment activation.

No causal or population claim follows from one ordered pair. The shorter trajectory may reflect treatment, stochasticity, cache state, or a different repair path. Future runs must first freeze a whole-tree baseline under the same sandbox and classify environment-inapplicable tests. They should then require both task success and no new failures relative to that baseline. This revised rule must be declared before new runs and cannot retroactively reclassify the present pair.

\subsection{The event graph is an operational audit object}

The graph improves auditability because failures have a typed locus. A policy denial belongs to a decision edge; a panic becomes an error result; compaction becomes a lifecycle event; and a missing result becomes a closure defect. This resembles an orchestration trace structurally~\cite{zhang2026orchestration}, but it is not a learned coordination policy or reward-bearing rollout. It is a projection of runtime state transitions that can be checked against local invariants.

Terminal closure matters independently of task success. A provider protocol may require each tool-use identifier to receive a result even when a tool is denied, cancelled, or fails. A useful final answer cannot retroactively repair an unpaired identifier, while a perfectly paired trace may still solve the wrong task. Runtime validity and task utility require separate measurements.

\subsection{Coverage is the central permission question}

Moving checks outside the prompt reduces dependence on the model following an instruction~\cite{mouzouni2026exploitation}, but it does not make authorization semantic. The gate evaluates the tool name, input, rules, scope hooks, and session mode available at that path. A harmful in-scope edit may be admitted; an authorized operation may be blocked when metadata or a classifier overestimates risk.

Coverage is path-dependent. The same external effect may be expressible through a built-in file tool, shell command, plugin, alias, or MCP service. Action-level permission work shows that aggregate task success can hide route-specific decisions~\cite{ji2026permission}. Metis resolves registered capabilities through a common registry and dispatch path, but external services and incorrectly declared tools remain trust boundaries. The computer-use service adds a second application gate; it does not inherit semantic intent from the main process.

\subsection{Continuity is not memory quality}

Context control prioritizes a provider-acceptable continuation: it prunes, snips, spills, summarizes, and repairs structural gaps. Memory research asks whether facts or procedures remain available and are used correctly later~\cite{sidik2026memtier,zhao2026npm}. A trace can be structurally valid after compaction yet omit evidence needed for a good decision. Current tests establish the former property only.

The same boundary applies to skills. A source-ranked \texttt{SKILL.md} file can be discovered and inserted without proving that the model will enact its procedure. Schema compilation and parametric-skill work target representation and behavioral uptake~\cite{sakizli2026tscg,zhao2026parametric}; Metis presently supplies deployment and visibility mechanisms rather than learned internalization.

\subsection{Delegation narrows authority but adds propagation paths}

A child loop receives a cloned gate and filtered registry, so delegation does not intentionally widen configured authority. This is weaker than process isolation. Children share the host and may share provider, network, budget, and spill state. Untrusted content can propagate from a parent's tool result into a child prompt, or return from a child into a later write. Typed events expose these transitions, but visibility is not a defense against indirect prompt injection~\cite{dingeto2026agentredbench}.

\section{Threats to Validity and Required Evaluation}
\label{sec:limitations}

\textbf{Construct validity.} Terminal events are log records, not independent observations. Forced serialization isolates a dispatcher mechanism but is not an alternative runtime. The permission oracle measures declared decisions without executing effects, the provider protocol is a single-Read marker task, and historical fail-to-pass oracles validate tasks rather than agent solutions. In the exploratory maintenance trace, task-local success is reported separately from the preregistered outcome that requires a clean whole-tree regression. The terms ``closure'' and ``continuation'' remain restricted to result identifiers and protocol structure.

\textbf{Internal validity.} The paired dispatcher design and alternating order reduce stable host and order effects, but filesystem caches, process scheduling, loopback transport, and the programmed HTTP delay remain part of that workload. The bootstrap interval quantifies pair variation, not bias outside this host. The exploratory maintenance trace has one seeded pair, with baseline executed first, so model stochasticity, cache/order effects, and the treatment are not separable. The child treatment removes gate and registry narrowing together, so their individual effects are unidentified. The permission cases were handcrafted and run once each.

\textbf{External validity.} The real-I/O probe covers local files, Git, loopback HTTP, and one write, not remote networks, large repositories, GUI operations, or competing runtimes. Five model conditions on one trivial protocol do not generalize to other events or tasks. Both frozen suites contain failures, and the companion coverage profile is partial. The one evaluable maintenance pair covers only panic containment, one model, one host, and one repetition; its task-local contrast and resource differences do not estimate maintenance effectiveness. Model availability failures remain excluded from protocol outcomes.

\textbf{Mechanism limits.} Scheduling correctness assumes accurate tool declarations; a hidden write labeled Safe can race. Terminal closure is not transactional rollback. Process or host loss may prevent repair. Permission checks can misclassify intent or miss equivalent effects. Compaction can remove decision-critical evidence. Computer-use gating has focus races and platform dependencies. Worktrees isolate repository writes, not the host or network.

Table~\ref{tab:futureprotocol} separates completed narrow evidence from the remaining studies needed to extend it.

\begin{table}[H]
\caption{\textbf{Studies required beyond the completed mechanism evidence.}}
\label{tab:futureprotocol}
\Description{Six remaining studies list their matched protocol, measurements, and the stronger claim each could support.}
\centering
\small
\renewcommand{\arraystretch}{1.05}
\setlength{\tabcolsep}{3.5pt}
\begin{tabularx}{\textwidth}{@{}B{28mm}L{57mm}Y@{}}
\toprule
Study & Matched protocol & Measurements and potential claim \\
\midrule
Executed permission routes & Fix intended state delta and policy; vary built-in, shell, plugin, alias, and MCP routes & Decisions plus safely sandboxed realized deltas could extend the current decision-only oracle \\
Indirect-output propagation & Fix task, integrations, and policy; vary clean/injected output and parent/child propagation & Attack success, event path, intervention, and overhead could establish read-to-write robustness \\
Context retention & Fix injected facts, procedures, and later queries; vary position, transformation, interruption, and spill availability & Protocol validity, recall, and decision consistency could separate structural from semantic retention \\
External recovery & Extend faults across process/host loss and remote services; add an explicit transactional design if rollback is claimed & State delta and recovery time could map a broader failure envelope; current partial mutation persists \\
Component child ablation & Hold registry or gate fixed while ablating the other over multiple authorized and unauthorized cases & Separate component effects and uncertainty could replace the current combined two-condition observation \\
Matched maintenance & Freeze the sandbox-specific whole-tree baseline, then repeat randomized pairs over four tasks while varying one engaged treatment & Task-macro success, no-new-regression rate, time, overhead, failures, and trace completeness could estimate a runtime contribution \\
\bottomrule
\end{tabularx}
\end{table}

\section{Conclusion}

Metis converts model-proposed tool calls into typed runtime events with explicit permission, scheduling, terminal-result, and lifecycle edges. Thirty matched real-I/O pairs supported the four-class dispatcher on one frozen implementation. Fault, child-boundary, permission-route, and model-protocol studies then mapped both supported mechanisms and unresolved boundaries.

Together, these results show that runtime behavior can be inspected independently of the model policy that proposes a call. The evidence remains mechanism-specific and does not establish product-level safety or cross-runtime superiority. The maintenance trace is retained as treatment-activation evidence, while repeated randomized tasks with a predeclared whole-tree baseline remain necessary for an effectiveness claim.

\section*{Artifact Availability}

A de-identified replication artifact is being curated for public release and is not part of this preprint version. It will contain normalized study records, task definitions, exclusion decisions, and deterministic analysis scripts. No human-participant or private user dataset was used.

\begingroup
\small
\setlength{\bibsep}{2pt plus 0.2ex}
\bibliographystyle{ACM-Reference-Format}
\bibliography{refs}
\endgroup

\end{document}